\documentclass[aps,prb,showpacs,twocolumn]{revtex4}
\usepackage{amssymb}
\usepackage{amsmath}
\usepackage{graphicx}
\usepackage{dsfont}
\usepackage{times}

\begin{document}

\title{Identifying Product Order with Restricted Boltzmann Machines}
\author{Wen-Jia Rao,$^{1,2}$ Zhenyu Li,$^1$ Qiong Zhu,$^3$ Mingxing Luo,$^1$ and
Xin Wan$^{1,2}$}
\affiliation{$^1$Zhejiang Institute of Modern Physics, Zhejiang University, Hangzhou
310027, China}
\affiliation{$^2$Collaborative Innovation Center of Advanced Microstructures, Nanjing
210093, China}
\affiliation{$^3$International Center for Quantum Materials, Peking University, Beijing
100871, China}
\date{\today}

\begin{abstract}
Unsupervised machine learning via a restricted Boltzmann machine is
an useful tool in distinguishing an ordered phase from a disordered phase.
Here we study its application on the two-dimensional Ashkin-Teller model,
which features a partially ordered product phase.
We train the neural network with spin
configuration data generated by Monte Carlo simulations and
show that distinct features of the product phase can be learned
from non-ergodic samples resulting from symmetry breaking.
Careful analysis of the weight matrices inspires us to define a
nontrivial machine-learning motivated quantity of the product form,
which resembles the conventional product order parameter.
\end{abstract}

\maketitle

\section{\protect\bigskip Introduction}

One of the central tasks of condensed matter physics is to identified phases
and phase transitions.
The conventional approach introduces the concept of order parameter,
which is a quantity that vanishes in a disordered phase but nonzero
in the adjacent ordered phase.
The value of the order parameter can be used to identify the transition
between the two phases, and plays a central role in the Landau theory
of phase transitions.
The fluctuations of the order parameter, which is closely related to
its dimension and symmetry, are crucial in understanding
the corresponding phase transition.
In practice, the choice of the order parameter is not unique,
but may not be obvious sometimes, such as in metal-insulator transitions.

Recent developments in machine learning (ML)~\cite{Goodfellow} have
found growing applications in the study of phases and phase
transitions.~\cite{Torlai16,Carrasquilla17,Tanaka16,Nieuwenburg17,Liu17,
Li17,Schindler17,YiZhang16,YiZhang17,Chng16,chng17,Wang16,ohtsuki16}
In these studies computer algorithms identify patterns in the
configurations of physical systems just
as they recognize images in the field of artificial intelligence.
Utilizing the knowledge learnt from data analysis,
one can also use machine learning schemes to improve existing numerical
algorithms.~\cite{JLiu16,JLiu17,LHuang17,LHuang172,Xu16,Nagai17,Broecker16}
Despite all these successful applications, the power and limitation
of ML remains to be understood.
It is tempting to seek connections to fundamental concepts in physics,
such as symmetry, locality, and renormalization group.\cite{Mehta14,Lin16,Maciej17}

Among various machine learning schemes, the restricted Boltzmann machine (RBM)
bears the closest analogy to physical systems.
The joint probability distribution of the model is a Boltzmann
distribution whose energy functional describes the couplings
between a visible layer and a hidden layer of spins.
RBMs can be used as generative models in machine vision or language
processing to extract high-level features.
Obvious applications in physics include representing probability
distributions, such as the Boltzmann distribution in calculating
partition functions, the probability density of wave functions,
or complex wave functions themselves.
For example, Torlai and Melko\cite{Torlai16} applied RBMs to study the thermodynamics of classical Ising models.
Amin {\it et al.}~\cite{Amin16} further generalized the RBM approach to study quantum models.
Carleo and Troyer~\cite{Carleo17} demonstrated that
using RBM representations as variational wave functions
one can approach even lower ground state energies than
with methods based on tensor networks.
In models of stablizer codes, RBMs can be shown to represent exact
ground states.~\cite{Deng16,Deng17}
The connection between the representative power of RBM and
tensor network states has been explored.\cite{Huang17,Chen17,Gao17}
Morningstar and Melko~\cite{morningstar17} found that the shallow RBM is more efficient
than its deep generalizations in representing physical probability
distributions, at least for Ising systems near criticality.

In this paper we apply an unsupervised learning with RBMs
to the $N_c$-color Ashkin-Teller (AT) model on a square lattice.
One motivation is that the AT model features a fully disordered
paramagnetic phase and a partially ordered product phase.
The two bear strong similarities in spin configurations
of any single color and both possess a large entropy.
It is, therefore, an interesting question whether the RBM
can distinguish the two phases.
In addition, the conventional order parameter of the
product phase is constructed by the product of two spins
of different color on the same lattice site.
The operator, however, is not directly present in the
energy functional of the RBM.
One wonders how ML can capture the product
order with the nontrivial order parameter.
To answer these questions, we organize our paper as follows.
In Sec.~\ref{sec:model} we describe the physical AT model and
the RBM neural network model. We explain how we train the RBM
with spin configurations from Monte Carlo simulations.
We discuss the optimal number of hidden nodes in
Sec.~\ref{sec:optimum_hidden}.
The results of the training are presented in Sec.~\ref{sec:results}.
We conclude with discussions of the results and on
possible further directions in Sec.~\ref{sec:discussion}.

\section{Model and Method}
\label{sec:model}

\begin{figure}[t]
\centering
\includegraphics[width=8.6cm,height=7cm]{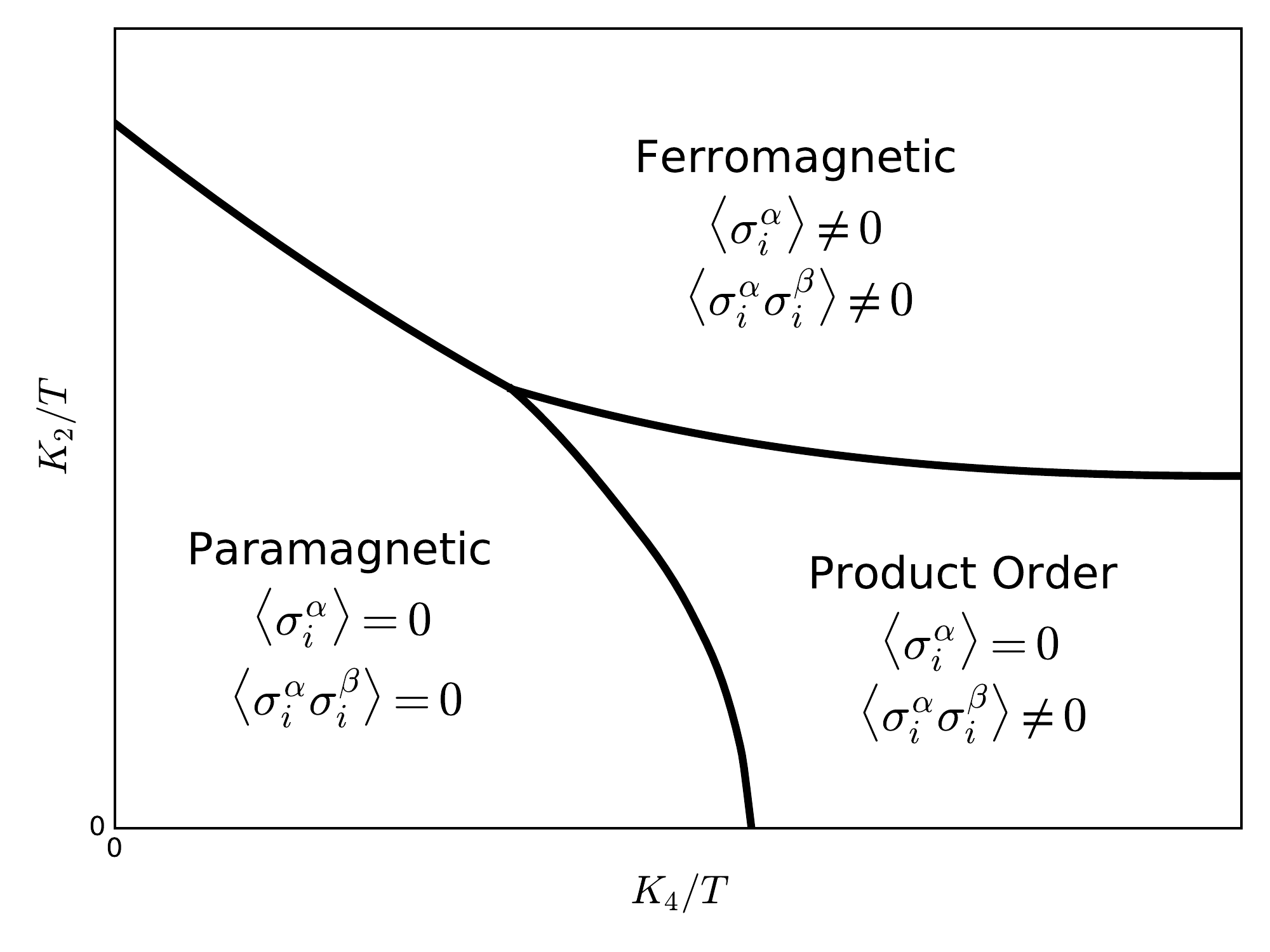}
\caption{Illustration of the phase diagram of the $N_{c}$-color AT model.
The intercept along the $K_2$ axis is $K_{2} / T = \ln (1+\sqrt{2}) / 2 \approx 0.4407$,
which is the 2D square-lattice Ising critical point and independent of
the number of colors $N_c$.
The intercept along the $K_4$ axis is $N_c$-dependent.
For the two-color AT model the tricritical point is self-dual at ($%
K_{t}/T$, $K_{t}/T$), where $K_{t}/T=(\ln 3)/4\approx 0.2746$.}
\label{fig:PD}
\end{figure}

We consider the homogeneous $N_{c}$-color AT model on a
two-dimensional (2D) square lattice. In the AT model each lattice site hosts $%
N_{c}$ colors (or species) of Ising spins, which are coupled through the
Hamiltonian
\begin{equation}
\label{eq:ATmodel}
H_{\mathrm{AT}}=-K_{2}\sum_{\left\langle i,j\right\rangle }\sum_{\alpha
=1}^{N_{c}}\sigma _{i}^{\alpha }\sigma _{j}^{\alpha
}-K_{4}\sum_{\left\langle i,j\right\rangle }\sum_{\alpha <\beta }\sigma
_{i}^{\alpha }\sigma _{i}^{\beta }\sigma _{j}^{\alpha }\sigma _{j}^{\beta }.
\end{equation}%
where $i$ and $j$ are lattice site indices while $\alpha$ and $\beta$ are color indices.
We restrict ourselves to the parameter space with $K_{2}>0$ and $K_{4}>0$. The first
term describes $N_{c}$ independent copies of the 2D nearest-neighbor Ising
models, while the second term couples different species with
nearest-neighbor four-spin interaction. The 2D square-lattice Ising model
has a continuous phase transition from ordered (ferromagnetic) to disordered (paramagnetic)
phases at $K_{2}/T = \ln (1+\sqrt{2}) / 2 \approx 0.4407$, which can be
characterized by a local order parameter $\langle \sigma _{i}\rangle $. For
large $K_{2}$ energy dominates and the Ising system is in the ferromagnetic
phase, while for small $K_{2}$ entropy dominates and the system is
disordered. In the presence of the four-spin interaction, a new phase
emerges when entropy competes favorably to the two-spin interaction energy,
but not to the four-spin interaction energy. Accordingly, the phase can be
characterized by a non-zero product order parameter $\langle O_{i}^{\alpha
\beta }\rangle \equiv \langle \sigma _{i}^{\alpha }\sigma _{i}^{\beta
}\rangle \neq 0$, even though there is no order among individual species,
i.e. $\langle \sigma _{i}^{\alpha }\rangle =0$. This partially ordered phase
is known as the product phase.
The partially ordered product phase still possesses a large spin-orientation
entropy [$O(N \ln2)$], which is of the same order as that of the disordered
phase. Based on the spin configuration of any single color,
one cannot distinguish the product phase from the paramagnetic phase.
The generic phase diagram for the homogeneous
$N_{c}$-color AT model is illustrated in Fig.~\ref{fig:PD}. In particular,
the AT model with $N_{c}=2$ can be solved exactly at the self-dual tricritical point
($K_2$, $K_4$) = ($K_{t}$, $K_{t}$),
where $K_{t}/T=(\ln 3)/4\approx 0.2746$.\cite{Baxter}
In the neural network context, the AT spin glass has been considered
as a generalization of the Hopfield model for the storage
and retrieval of embedded patterns.~\cite{bolle98}

\begin{figure}[t]
\centering
\includegraphics[width=8.6cm]{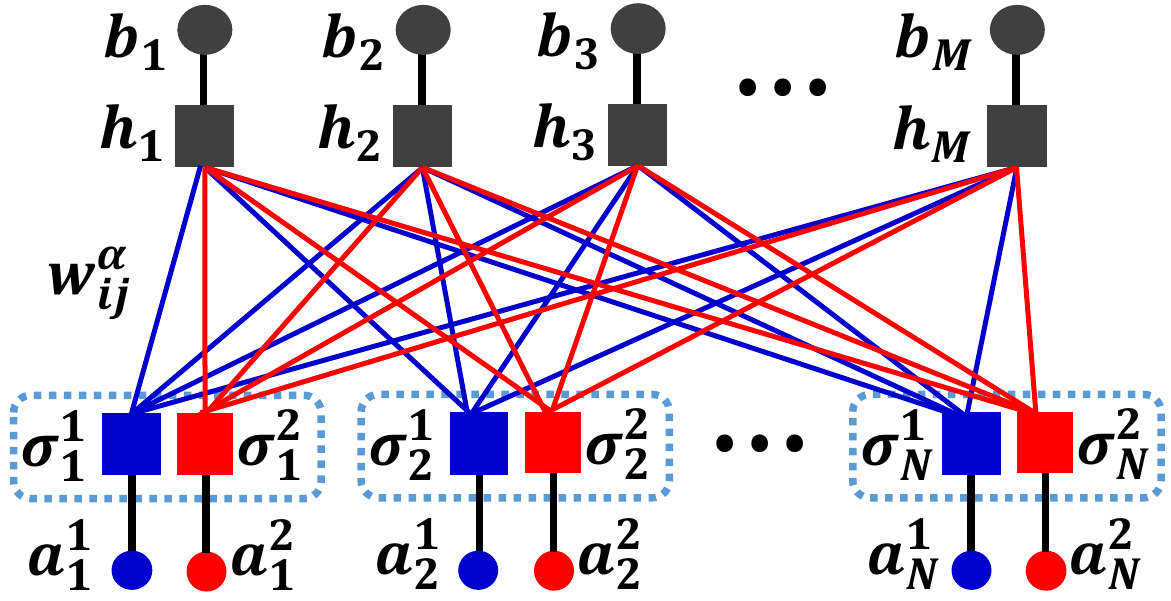} % Here is how to import EPS art
\caption{Illustration of the RBM for the two-color AT model.
The dotted boxes represent $N$ lattice sites, on each of which there are two
Ising spins $\sigma_i^{\alpha}$ with different color $\alpha$
(blue and red squares for $\alpha = 1,2$).
They comprise the visible layer.
The hidden layer consists of $M$ hidden Ising spins $h_j$ (dark grey boxes).
There are no intralayer couplings, but spins in the two layers are coupled via
edges $w_{ij}^{\alpha }$. External field $%
a_{i}^{\alpha }$ (blue and red dots) and $b_{j}$ (dark grey dots) are introduced
for the visible and hidden layers, respectively.
This graph structure can be easily generalized to the $N_c$-color
AT model with $N_c > 2$.}
\label{fig:RBM}
\end{figure}

The RBM for the AT model is illustrated in Fig.~\ref{fig:RBM}.
$N$ lattice sites, each with $N_c$ physical Ising spins (or nodes) $\sigma _{i}^{\alpha }$,
form a visible layer, while an additional $M$ Ising spins (or nodes) $h_{j}$ form a
hidden layer. Local fields $a_{i}^{\alpha }$ and $b_{j}$ are applied to the
visible and hidden nodes, respectively. In a RBM couplings $%
w_{ij}^{\alpha }$ exist as edges only between two nodes in different layers,
hence the modifier restricted.
We emphasize that we color-code the $N_c$ visible spin species,
as well as the corresponding edges, in Fig.~\ref{fig:RBM}
by different colors (blue for $\alpha = 1$, red for $\alpha = 2$,
green for $\alpha = 3$, etc.) and will present our results
with such a color scheme.
Mathematically, this graph describes
a joint probability distribution%
\begin{equation}
p_{\lambda }\left( \mathbf{\sigma ,h}\right) =
\frac{e^{-E_{\lambda}(\mathbf{\sigma ,h})}} {Z_{\lambda }},
\end{equation}
where
\begin{equation}
E_{\lambda}(\mathbf{\sigma ,h}) = -\sum_{i,\alpha }a_{i}^{\alpha }\sigma _{i}^{\alpha
}-\sum_{j}b_{j}h_{j}-\sum_{i,\alpha ,j}\sigma _{i}^{\alpha }w_{ij}^{\alpha
}h_{j}.
\end{equation}%
Here, the subscript $\lambda $ stands for the collection of RBM parameters $%
\left \{ a,b,w\right \} $, and $Z_{\lambda }$ is the normalizing partition
function. The probability distribution for the visible nodes is then%
\begin{equation}
p_{\lambda }\left( \mathbf{\sigma }\right) =\sum_{\mathbf{h}}p_{\lambda }\left( \mathbf{\sigma },\mathbf{h}\right) \equiv \frac{1}{%
Z_{\lambda }}\exp \left( -E_{\lambda }\left( \mathbf{\sigma }\right) \right)
\end{equation}%
where the model energy functional
\begin{equation}
E_{\lambda }\left( \mathbf{\sigma }\right) =-\sum_{i,\alpha }a_{i}^{\alpha
}\sigma _{i}^{\alpha }-\sum_{j}\log \left(1+e^{b'_j} \right)
\label{equ:RBM}
\end{equation}%
and the effective field for the $j$th hidden spin
\begin{equation}
b'_j = b_{j} + \sum_{i,\alpha }\sigma _{i}^{\alpha }w_{ij}^{\alpha }.
\end{equation}
It has been proved\cite{Bengio08,Montufar11} that such a distribution can
describe an arbitrary probability distribution provided that the number of
hidden nodes is sufficiently large. In the present study, we feed spin
configurations of the AT model to the RBM with the objective to train the
RBM distribution $p_{\lambda }\left( \mathbf{\sigma }\right) $ to represent
the Boltzmann distribution of the classical Hamiltonian as closely as
possible; in other words, we approximate the original Hamiltonian with the
effective RBM energy functional $E_{\lambda }\left( \mathbf{\sigma }\right) $
in Eq.~\ref{equ:RBM}. Previously, the RBM approach has been applied to the
classical Ising model in both one and two dimensions.\cite{Torlai16} For the
AT model, a visible spins contain an additional color index and, therefore,
the coupling between the visible spin and any hidden node also contains the
corresponding color index. As we will show later, the correlation between
the couplings of a hidden node to a pair of visible spins with the same site
index but different color indices is crucial for the understanding of the
partially ordered product phase.

To obtain an optimized set of parameters $\lambda \equiv \left( a,b,w\right)
$, we define the standard KL divergence%
\begin{equation}
KL\left( \lambda \right) =\sum_{\mathbf{\sigma }}p_{B}\left( \mathbf{\sigma }%
\right) \log \frac{p_{B}\left( \mathbf{\sigma }\right) }{p_{\lambda }\left(
\mathbf{\sigma }\right) }
\end{equation}%
as a cost function to judge how close $p_{\lambda }(\sigma )$ is to the
Boltzmann distribution function
\begin{equation}
p_{B}(\mathbf{\sigma },T)={\frac{1}{Z_{\mathrm{AT}}}}e^{-H_{\mathrm{AT}}(%
\mathbf{\sigma })/T},
\end{equation}%
where the partition function
\begin{equation}
Z_{\mathrm{AT}}=\mathrm{Tr}\left[ e^{-H_{\mathrm{AT}}(\mathbf{\sigma })/T}%
\right] .
\end{equation}%
It can be shown that $KL\left( \lambda \right) \geq 0$; the equality holds
only when $p_{B}$ coincides with $p_{\lambda }$. Hence, \emph{training} is
an optimization procedure that updates parameters $\lambda $ by reducing the
KL divergence.

We follow the standard machine learning procedure to train the RBM machine
with the spin configurations of the AT model generated by Monte Carlo
simulation. If we denote the data set by $D \equiv \left\{ \mathbf{\sigma }%
^{\left( 1\right) },\mathbf{\sigma }^{\left( 2\right) }, \dots, \mathbf{\sigma }%
^{\left( D\right) }\right\} $, we thus replace the Boltzmann distribution by
the probability distribution $p_{data}\left( \mathbf{\sigma }\right) =\frac{1%
}{D}\sum_{\mathbf{\sigma }^{\prime }\in D}\delta \left( \mathbf{\sigma },%
\mathbf{\sigma }^{\prime }\right) $. The KL divergence is then simply%
\begin{equation}
KL\left( \lambda \right) =-\frac{1}{D}\sum_{\mathbf{\sigma }^{\prime }\in
D}\log p_{\lambda }\left( \mathbf{\sigma }^{\prime }\right) -H\left( p_{%
\mathrm{data}}\right)  \label{equ:KL}
\end{equation}%
where%
\begin{equation}
H\left( p_{\mathrm{data}}\right) =-\frac{1}{D}\sum_{\mathbf{\sigma }^{\prime
}\in D}\log \left( \frac{1}{D}\sum_{\mathbf{\sigma }^{\prime \prime }\in
D}\delta \left( \mathbf{\sigma }^{\prime },\mathbf{\sigma }^{\prime \prime
}\right) \right)
\end{equation}%
The separation is convenient, because only the first term in the KL
divergence depends on $\lambda$ and needs to be updated during training,
while the second term $H\left( p_{data}\right) $, or the entropy of the data
set, needs to be computed only once.

For the optimization of the KL divergence in the presence of a large data
set, we adopt the stochastic gradient descent (SGD) approach. The approach
involves repeated calculations of the gradient of the log-likelihood $\log
p_{\lambda }\left( \mathbf{\sigma }\right) $ with respect to any of the
parameters $\lambda_k \equiv a_{i}^{\alpha }$, $b_{j}$, or $w_{ij}^{\alpha
}$,
\begin{equation}
\nabla _{\lambda _{k}}\log p_{\lambda }\left( \mathbf{\sigma }\right)
=-\nabla _{\lambda _{k}}E_{\lambda }\left( \mathbf{\sigma }\right) +
\sum_{\mathbf{\sigma' }} \frac{e^{-E_{\lambda } \left( \mathbf{\sigma' }%
\right)}}{Z_{\lambda }}  \nabla _{\lambda _{k}}E_{\lambda }\left( \mathbf{\sigma' }\right) .
\end{equation}%
Note that the evaluation of the normalization $Z_{\lambda }$ in the second
term involves a summation over exponentially large number of configurations,
hence impossible for practical calculations of the log-likelihood gradient.
Instead, we adopt the $k$-step contrastive divergence (CD$_{k}$),\cite{CDK}
which approximates the gradient locally around the training data.

With these methods and approximations, we start from a randomly chosen
initial parameters $\lambda ^{\left( 0\right) }$ and update through%
\begin{equation}
\lambda ^{\left( n+1\right) }=\lambda ^{\left( n\right) }-\eta \nabla
_{\lambda }KL\left( \lambda ^{\left( n\right) }\right)
\end{equation}%
where the coefficient $\eta$ is the learning rate, whose value needs to be
carefully chosen to balance the speed to explore the parameter space and the
stability. After sufficiently long steps, we terminate the training and
explore the resulting parameter set $\lambda$ and try to associate certain
patterns to various phases, in particular, to the product phase.

\section{Machine-Learning Results}
\label{sec:results}

In the following RBM study we use $200$ hidden nodes and
initiate $\lambda^{(0)}$ with uniform distribution within
$\left[ -0.03,0.03\right]$.
For the training, we choose the CD$_{20}$ approximation and
a fixed learning rate $\eta =0.03$.
We have checked that the qualitative results we present below
are robust against the variation of these super parameters within
reasonable range.

\subsection{Similarities between the product and paramagnetic phases}
\label{sec:results.puzzle}

We begin our discussion on the identification of the product
phase of the two-color AT model in an $8 \times 8$ lattice.
We first perform Monte Carlo sampling to generate a data set of
$10^{5}$ configurations for the following two parameter sets:
(i) $K_{2}/T=0.1$ and $K_{4}/T=0.1$ in the paramagnetic phase, and
(ii) $K_{2}/T=0.1$ and $K_{4}/T=1$ in the product phase.
We choose to compare the two cases
both with $\langle \sigma _{i}^{\alpha }\rangle =0$.
The parameters are chosen that the states are sufficiently far away
from the phase boundary, but we have ensured that the ergodicity
in (ii) is not broken in our simulation.
We then feed the data into the RBM to learn the Boltzmann
distribution of the AT model.

\begin{figure}
\centering
\includegraphics[width=8.6cm]{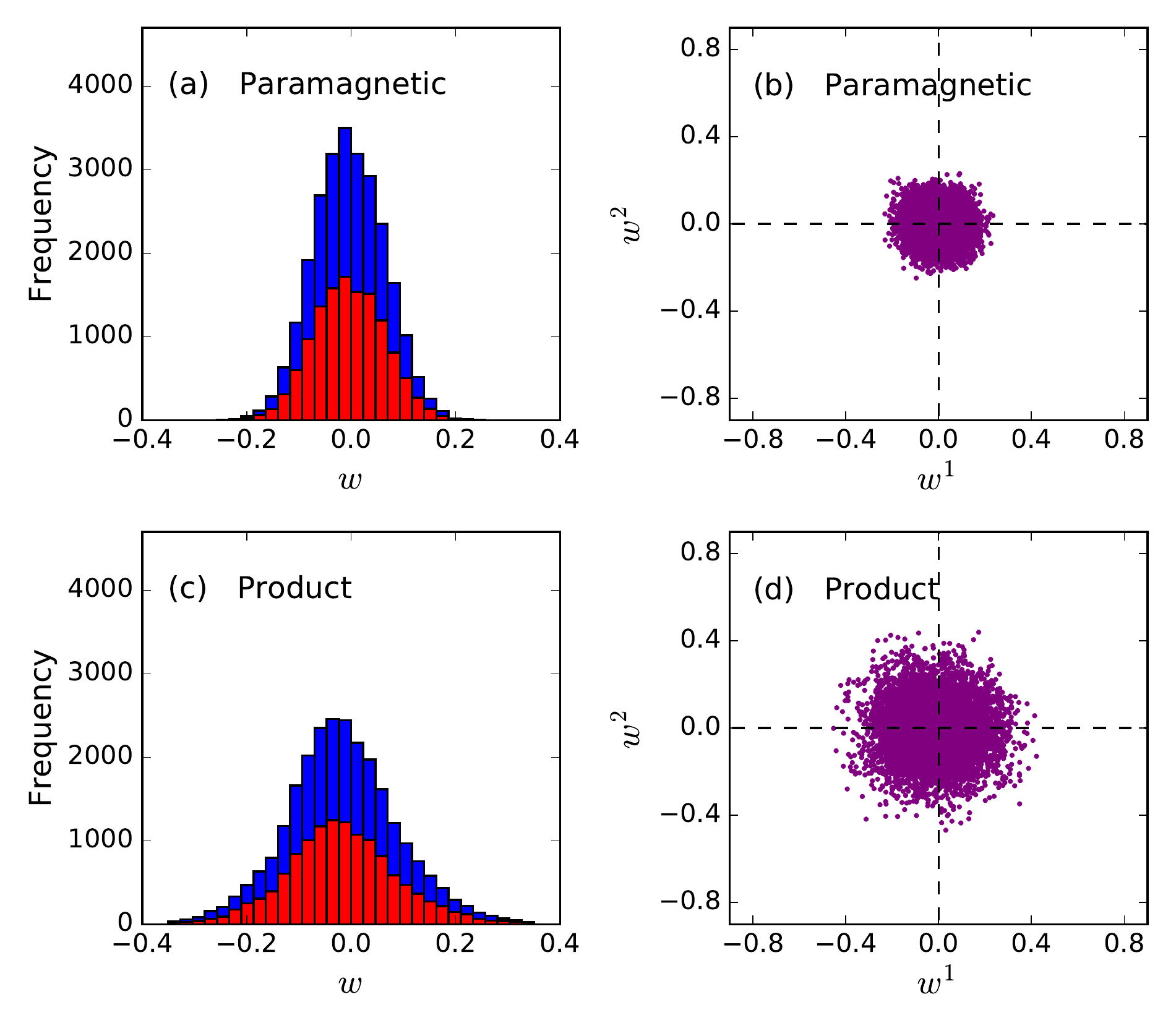}
% Here is how to import EPS art
\caption{Histogram of the coupling coefficients $w_{ij}^{\alpha }$ of
the two-color AT model on a 8$\times$8 square lattice for
(a) $K_{2}/T=0.1$ and $K_{4}/T=0.1$ in the paramagnetic phase, and
(c) $K_{2}/T=0.1$ and $K_{4}/T=1$ in the product phase.
They are color coded such that blue stands
for color index $\alpha = 1$ and red for $\alpha = 2$.
Comparison of the corresponding coupling coefficients
$w_{ij}^{1}$ versus $w_{ij}^{2}$ for the two colors
are shown in (b) for the paramagnetic phase and
in (d) for the product phase.
}
\label{fig:para_product}
\end{figure}

In Figure~\ref{fig:para_product}(a) and (b) we compare
the histograms of the interlayer coupling coefficients $w_{ij}^{\alpha }$
for the two parameter sets in the paramagnetic and product phases.
We break the histograms of the couplings into two colors
according to their connections to the visible spins with
the corresponding colors (blue for $\alpha = 1$
and red for $\alpha = 2$).
The histograms show a similar bell-shape in both paramagnetic
and product phases. After fitting the histograms by a Gaussian
distribution function $p(w)=p_0 e^{-(w-\mu)^2/2\zeta^2}$,
we find $\mu=-0.0049$ and $\zeta=0.069$ for the paramagnetic
phase and $\mu=-0.021$ and $\zeta=0.096$ for the product phase.
In both cases the deviation of the peak from the origin is insignificant.
The width of the two bells are similar; in the product phase the width
is about 39\% larger.
In the earlier RBM learning of the 2D Ising model, Torlai and Melko~\cite%
{Torlai16} observed a sharp contrast in the histogram of the couplings
between the ferromagnetic and paramagnetic phases.
This is understandable because strong interlayer couplings are necessary
to generate the long-range spin correlation in the ferromagnetic phase,
while weak interlayer couplings imply that the visible spins behave
like independent entities.
We generated spin configurations in the ferromagnetic phase
of the AT model by the Wolff algorithm.
Due to the presence of the color index, more hidden nodes are needed
in the training than those in the Ising case for the same lattice size.
Qualitatively, we confirm that the width of the coupling strength
distribution can be regarded as an order parameter
to distinguish the ferromagnetic phase and
the paramagnetic phase.
Unfortunately, as we showed above, this would be an unreliable
indicator for identifying the product phase.

The existence of the product order also prompts us to explore the correlation
between the couplings connecting the same hidden spin and the two
visible spins with different colors on the same lattice site.
We plot the corresponding $w_{ij}^{1}$ versus $w_{ij}^{2}$
in Fig.~\ref{fig:para_product}(b)
for the paramagnetic phase and
in Fig.~\ref{fig:para_product}(d) for the product phase.
In both cases, the data scatters around the origin with no significant
enhancement in any of the four quadrants.
Once again, we cannot distinguish the paramagnetic and the product
phases according to the RBM parameters.
Is it, then, possible to define an machine-learning motivated order
parameter based on the RBM parameters for the product phase?

\subsection{Non-ergodicity and the product order}
\label{sec:results.nonergodic}

Let us digress a moment to the ferromagnetic case.
The conventional order parameter is the magnetization,
whose emergence is related to the breaking of the $Z_2$ symmetry
of Ising spins.
Below the critical temperature, magnetization can take on two signs.
In the thermodynamic limit, ergodicity is broken and
the magnetization is either positive or negative.
In a finite system, however, ergodicity may not be broken in a Monte
Carlo simulation, especially with the implementation of
various cluster algorithms.
Whether the ergodicity is broken or not will not affect the study of
phases and phase transitions as long as we take proper care of
the sign of the order parameter.
How important, then, is the ergodicity broken in the RBM study?
With these considerations in mind, we now return to the product phase
to show that ergodicity breaking indeed holds a crucial role
in the identification of an appropriate order parameter.

\begin{figure}
\centering
\includegraphics[width=8.6cm]{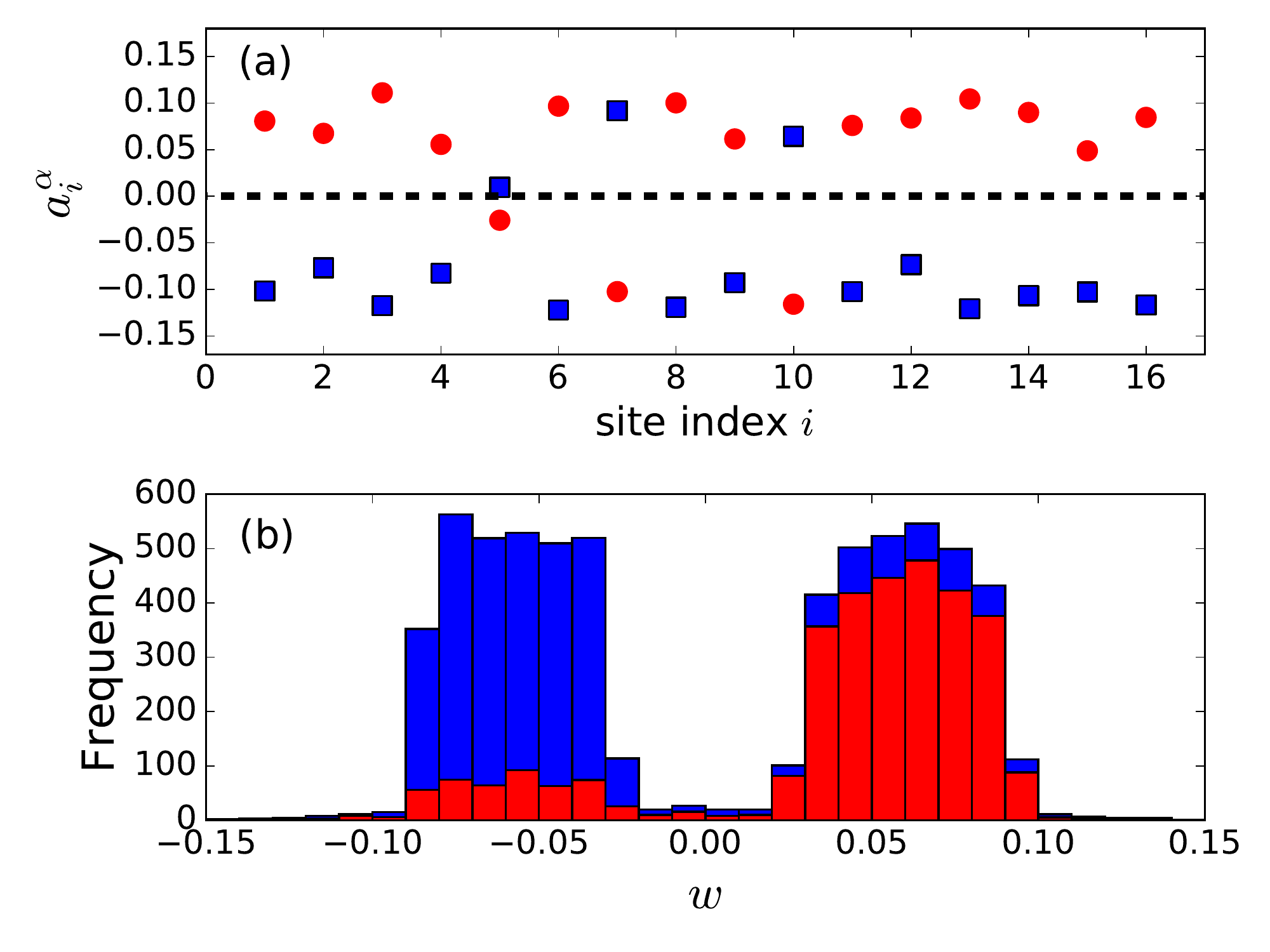}
% Here is how to import EPS art
\caption{RBM parameters obtained from the training of the two-color AT model
on a 4$\times$4 square lattice with $K_{2}/T=0.1$ and $K_{4}/T=2$ in the product phase.
(a) The local fields $a_{i}^{1}$ and $a_{i}^{2}$ for the two-color Ising spins
on any site $i$ are of opposite signs.
(b) The histogram of the coupling coefficients $w_{ij}^{\alpha }$ exhibits two lumps,
one on each side of the origin.
The figure is color coded such that blue stands for color index $\alpha = 1$ and
red for $\alpha = 2$.}
\label{fig:2Color-Product}
\end{figure}

In the following we train the RBM with ergodicity-breaking data
in a $4 \times 4$ lattice for $K_{2}/T=0.1$ and $K_{4}/T=2.0$.
We note that deeper in the product phase spin configurations
generated by Metropolis sampling are ergodicity broken,
unless the simulations time is sufficiently long.
Figure~\ref{fig:2Color-Product} shows the local fields $a_{i}^{\alpha }$ and
the histogram of the coupling coefficients $w_{ij}^{\alpha }$
in the product phase. The local fields
for color-1 (blue) spins are predominantly negative (13 out of 16), while
those for color-2 (red) are predominantly positive (also 13 out of 16).
Furthermore, they are so correlated that $a_{i}^{1}a_{i}^{2}<0$ for every
lattice site $i$. Intriguingly, the histogram of $w_{ij}$ develops two broad
but separate lumps on the two sides of the origin. We intentionally plot the
histogram of $w_{ij}^{\alpha }$ for individual colors. Obviously, the two
lumps do not correspond to the two colors, even though the majority of $%
w_{ij}^{\alpha }$ connecting blue spins clusters with negative values, while
the majority of those connecting red spins clusters with positive values.
The broken color symmetry among $a_{i}^{\alpha }$ or among $w_{ij}^{\alpha }$
motivates us to further explore the product of these parameters. Figure~\ref%
{fig:2C-Product-WW-AW} shows the histogram of $w_{ij}^{1}w_{ij}^{2}$ and $%
a_{i}^{\alpha }w_{ij}^{\alpha }$. We find that \emph{almost all} (above
99.9\%) $w_{ij}^{1}w_{ij}^{2}$ are negative and 96.8\% of $a_{i}^{\alpha
}w_{ij}^{\alpha }$ are positive, which means that the signs of $%
a_{i}^{\alpha }$ and $w_{ij}^{\alpha }$ associating with visible spins on
the same lattice site $i$ are well correlated. This again is not hard to
understand if one explores the effective energy in Eq.~\ref{equ:RBM}. In the
product phase with $\left\langle \sigma _{i}^{1}\sigma _{i}^{2}\right\rangle
<0$, negative $w_{ij}^{1}w_{ij}^{2}$ can lead to negatively large $%
\sum_{i,\alpha }\sigma _{i}^{\alpha }w_{ij}^{\alpha }$, hence the sum in the
second term on the righthand side of Eq.~\ref{equ:RBM} is positively large.
Positive $a_{i}^{\alpha }w_{ij}^{\alpha }$ also imply that $a_{i}^{\alpha
}\sigma _{i}^{\alpha }$ in the first term are negative. As a
result, the effective energy is large but negative. In other words,
configurations with $\sigma _{i}^{1}\sigma _{i}^{2}<0$ have larger weight
than those with $\sigma _{i}^{1}\sigma _{i}^{2}>0$, hence the product order
develops. We would like to
point out that an independent sampling and the consequent training may as
well result in the dominance of positive $w_{ij}^{1}w_{ij}^{2}$, which
corresponds to the product phase with $\left\langle \sigma _{i}^{1}\sigma
_{i}^{2}\right\rangle >0$.

Interestingly, even though more than 99.9\% of the product $%
w_{ij}^{1}w_{ij}^{2}$ are negative, only 15.60\% of $w_{ij}^{1}$ are
positive. Neither is the latter percentage close to 0 or 1 as the
translational invariance may suggest, nor is it approaching 0.5 in a random
fashion. In fact, the number is, as it should, close to the percentage of $%
a_{i}^{1}$ being negative (3 out of 16). It turns out this number is
training-dependent. We point out that in the product order phase there is an
emergent local symmetry: $\sigma _{i}^{1}\leftrightarrow \sigma _{i}^{2}$,
which leaves the order parameter invariant. The RBM can accommodate this
symmetry by the joint transformation: $a_{i}^{1}\leftrightarrow a_{i}^{2}$
and $w_{ij}^{1}\leftrightarrow w_{ij}^{2}$ for all $j$. As a result, the
polarization of $a_{i}^{\alpha }$ or $w_{ij}^{\alpha }$ cannot serve as an
indicator of the product phase. Instead, the polarization of the product $%
w_{ij}^{1}w_{ij}^{2}$ or $a_{i}^{\alpha }w_{ij}^{\alpha }$ can. We
can use, e.g, the product $w_{ij}^{1}w_{ij}^{2}$ to design a
machine-learning motivated order parameter
\begin{equation}
\Gamma \equiv {\frac{1}{{NM}}}\sum_{i=1}^{N}\sum_{j=1}^{M}\mathrm{sgn}\left(
w_{ij}^{1}w_{ij}^{2}\right) ,
\label{eqn:gamma}
\end{equation}
where sgn($x$) is the sign function.
In the product phase we expect $\vert \Gamma \vert \approx 1$, while in the disordered
paramagnetic phase $\Gamma =O(1/\sqrt{N})$ due to fluctuations.

\begin{figure}
\centering
\includegraphics[width=8.6cm]{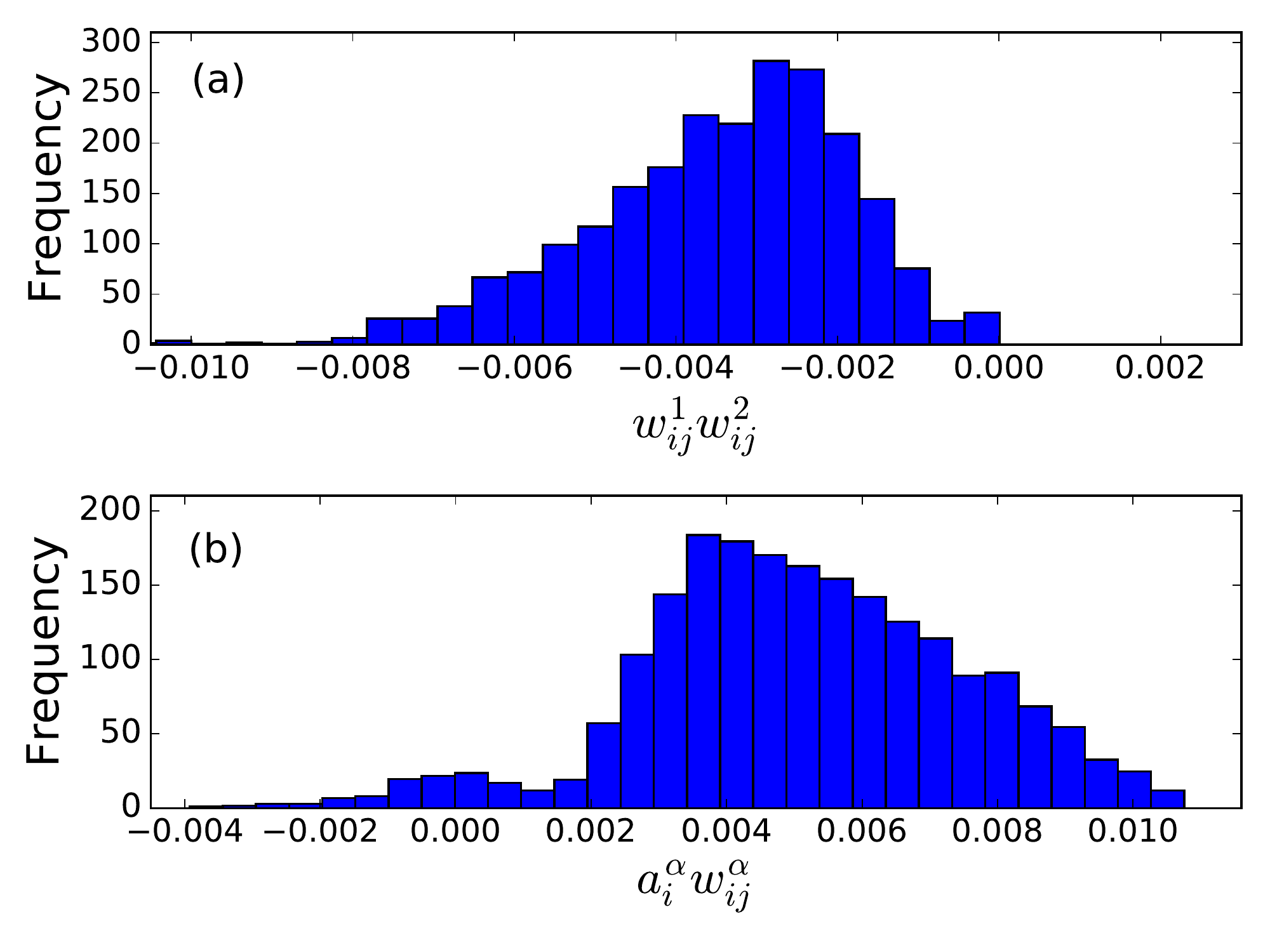}
% Here is how to import EPS art
\caption{Histograms of the products $w_{ij}^{1}w_{ij}^{2}$ and
$a_{i}^{\protect\alpha}w_{ij}^{\protect\alpha }$ resulting from
the training in the product phase, as illustrated in Fig.~\ref{fig:2Color-Product}.
(a) More than $99.9\%$ of $w_{ij}^{1}w_{ij}^{2}$ are negative.
(b) $96.8\%$ of $a_{i}^{\protect\alpha}w_{ij}^{\protect\alpha }$ are positive.}
\label{fig:2C-Product-WW-AW}
\end{figure}

To understand the evolution of the RBM parameters from a single Gaussian peak
to the two-lump structure when ergodicity is gradually broken, we vary
$K_{4}/T$ from $0.1$ to $2.0$ with fixed $K_{2}/T=0.1$.
The system undergoes a paramagnetic--product-order phase transition
in the two-color AT model around $K_{4_{C}}/T\simeq 0.4$ in
thermodynamic limit.\cite{Baxter}
We perform our training in the $L=4$ lattice
and keep using the Metropolis algorithm throughout the study.
For comparison, we choose the number of configurations for each $K_4/T$
to be $10^{5}$ and fix the sampling interval to be $\tau_{u}=1000$ MC sweeps.
We use identical super parameters for the RBM training and
fix the total number of training steps to be $4000$ before we
analyze the results.

\begin{figure}[t]
\centering
\includegraphics[width=8.6cm]{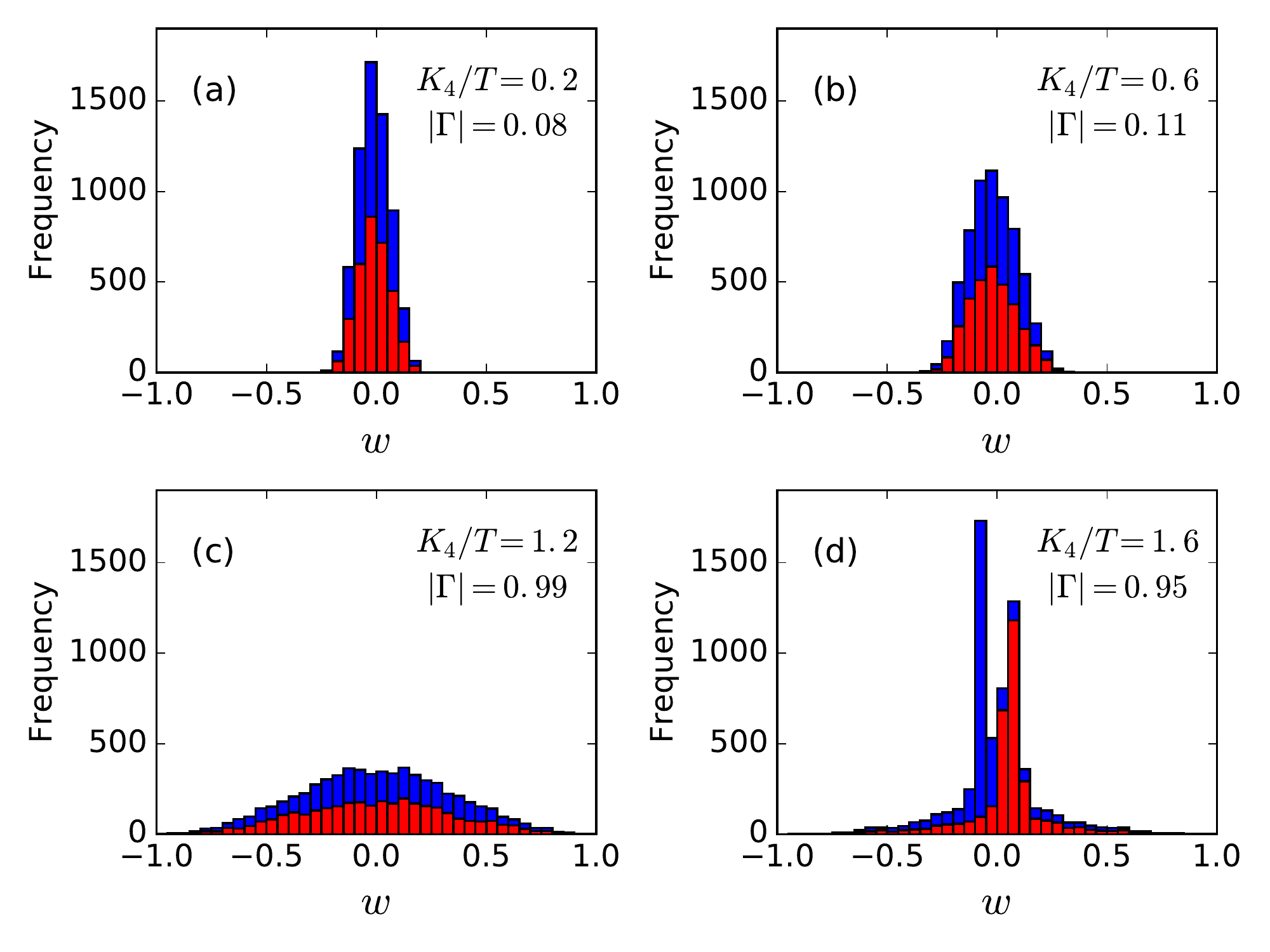}
\caption{The evolution of RBM parameters for the two-color AT
model in a $L=4\times 4$ lattice for various $K_{4}/T$
but fixed $K_{2}/T=0.1$.
(a) The histogram of $w_{ij}^{\alpha}$ at $K_{4}/T=0.2$
in the paramagnetic phase has a narrow Gaussian shape.
(b) In the product phase with $K_{4}/T=0.6$, the histogram
of $w_{ij}^{\alpha}$ remains Gaussian.
As in (a), the polarization in $w_{ij}^{1}w_{ij}^{2}$
is negligible.
(c) In the product phase with $K_{4}/T=1.2$,
the ergodicity in the product order is broken.
As a result, $\left\vert \Gamma \right\vert = 0.99$.
The Gaussian shape of the histogram of $w_{ij}^{\alpha}$
further widens.
(d) Deep in the product phase at $K_{4}/T=1.6$,
the ergodicity for spins of any individual color is also
broken within the fixed simulation time.
Two split but sharp peaks develops in the histogram of
$w_{ij}^{\alpha}$.
In all cases, data are color coded such that
blue stands for $\protect\alpha =1$ and red for $\protect%
\alpha =2$.}
\label{fig:L4_evolution}
\end{figure}

As $K_{4}/T$ increases, the behavior of the RBM parameters depends
crucially on the increase of the correlation time $\tau _{s}$ and $\tau _{p}$
of $\sigma_i^{\alpha}$ and $\sigma_i^{1}\sigma_i^{2}$, respectively.
We divide the parameter space into four regimes and exemplify
each region with a representative point in Fig. \ref{fig:L4_evolution}.
In Fig.~\ref{fig:L4_evolution}(a), $K_{4}/T = 0.2$.
The system is in the paramagnetic phase and the
Metropolis sampling is efficient.
As we present in Fig.~\ref{fig:para_product}(a) for a larger system,
the histogram of the coupling coefficients $w_{ij}^{\alpha}$
has a narrow Gaussian shape.
In Fig.~\ref{fig:L4_evolution}(b), $K_{4}/T = 0.6$. The system is in the product phase.
As we sample for a sufficiently long time, the configurations are ergodic.
As we discussed in Sec.~\ref{sec:results.puzzle},
the RBM parameters cannot be used to identify the product phase
qualitatively when ergodicity is preserved.
The histogram of $w_{ij}^{\alpha}$ is still Gaussian,
but with a wider peak.
We also find $\vert \Gamma \vert = 0.11$, which
is similar to $\vert \Gamma \vert = 0.08$ in (a),
so there is also no significant
polarization in $w_{ij}^{1}w_{ij}^{2}$.
In Fig.~\ref{fig:L4_evolution}(c), $K_{4}/T = 1.2$. The system is in product phase
and the ergodicity for the product operator is broken.
This case is very similar to the case to be discussed in
Sec.~\ref{sec:results.ergodic}, where the product symmetry
is explicitly broken by an external field.
In this case the histogram of $w_{ij}^{\alpha}$
is an even wider Gaussian peak,
but $w_{ij}^{1}w_{ij}^{2}$ is polarized,
as we find $\vert \Gamma \vert = 0.99$.
The sign of $\Gamma$ depends on the random polarization
direction due to the random importance sampling.
It is worth pointing out that the correlation time for
single color operators is still small, such that
ergodicity is still preserved within each color.
In Fig.~\ref{fig:L4_evolution}(d), $K_{4}/T\geq 1.6$.
Now deep in the product phase,
ergodicity is broken for both single-color spin flips and
the product order.
The resulting histogram of $w_{ij}^{\alpha}$
shows two split peaks, as in
Fig. \ref{fig:L4_evolution}(d), and the distributions for
$w_{ij}^{1}$ and $w_{ij}^{2}$ become different.
As $K_{4}/T$ further increases,
the long tails of the peaks disappear and
the two-lump structure develops as in
Fig.~\ref{fig:2Color-Product}(b).

The evolution from the paramagnetic phase to the ferromagnetic phase
with increasing $K_{2}/T$ is simpler,
as what matters is only the correlation time $\tau _{s}$ for
$\sigma_i^{\alpha}$ of any color index $\alpha$.
In the ferromagnetic phase with preserved ergodicity (e.g., achieved by
cluster updates), the histogram of $w_{ij}^{\alpha}$ exhibits
a broader distribution than that in the paramagnetic case.
When we explicitly break the ergodicity by single spin flips
in generating the spin configurations, the distribution shifts to
either the positive or the negative side, signaling the
$Z_2$ symmetry breaking in the thermodynamic limit.

\subsection{Breaking the product symmetry in ergodic samples}
\label{sec:results.ergodic}

To demonstrate the feasibility of $\Gamma$ as an order parameter for
the product order, we consider the following setup.
We start from the $8 \times 8$ lattice with $K_{2}/T=0.1$ and $K_{4}/T=1$
explored in Sec.~\ref{sec:results.puzzle}.
With preserved ergodicity in data from the simulation, we showed there
that the RBM parameters do not show significant difference from
those for a paramagnetic state.
In particular, there is no nontrivial color pattern in the parameter,
so one expects $\Gamma = 0$.
To be consistent, we also have a vanishing conventional product order parameter
\begin{equation}
\langle \sigma^1\sigma^2\rangle \equiv \frac{1}{N_s}\sum_{j=1}^{N_s}
\left(\frac{1}{N}\sum_{i=1}^{N}\sigma_i^1\sigma_i^2\right)_\text{{\it j}th configuration},
\label{equ:s1s2}
\end{equation}
where $N_s$ is the number of spin configurations fed into the RBM training.
We now introduce an ergodicity breaking term $H_{4}\sum_{i}\sigma _{i}^{1}\sigma _{i}^{2}$
to the original AT Hamiltonian Eq.~(\ref{eq:ATmodel}).
Depending on the sign of $H_4$, $\langle \sigma^1\sigma^2\rangle$ polarizes
accordingly.
For $-0.2 \leq H_4/T \leq 0.2$, we perform the RBM training to the spin configurations
obtained from sufficiently long simulations, such that ergodicity is not broken.

\begin{figure}[t]
\centering
\includegraphics[width=8.6cm]{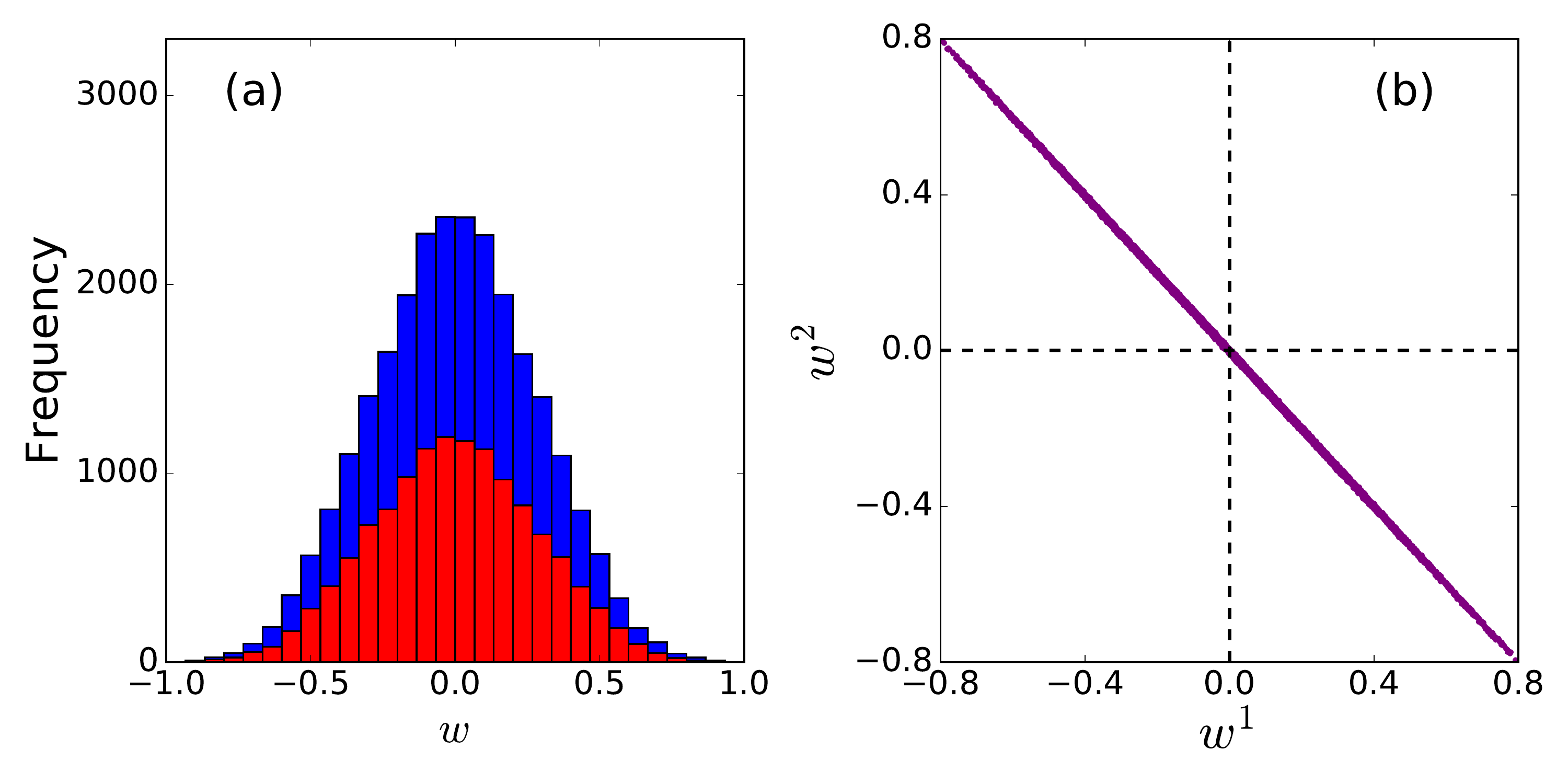}
\caption{(a) Histogram of the coupling coefficients $w_{ij}^{\alpha }$ of
the two-color AT model on a 8$\times$8 square lattice for
$K_{2}/T=0.1$ and $K_{4}/T=1$ in the product phase
with an additional symmetry breaking field
$H_4 \sum_{i}\sigma _{i}^{1}\sigma _{i}^{2}$, where $H_4 / T = 0.2$.
They are color coded such that blue stands
for color index $\alpha = 1$ and red for $\alpha = 2$.
(b) Comparison of the corresponding coupling coefficients
$w_{ij}^{1}$ versus $w_{ij}^{2}$. The straight line with slope
-1 is consistent with $\langle \sigma_i^{1}\sigma_i^{2} \rangle=-1$
for sufficiently large $H_4 > 0$.}
\label{fig:L8Prod_h4}
\end{figure}

Figure~\ref{fig:L8Prod_h4}(a) shows the training result for $H_4 / T =0.2$, at which the spin configurations
are totally polarized to $\langle \sigma_i^{1}\sigma_i^{2} \rangle=-1$
while spins of individual colors remain disordered.
Interestingly, however, the histogram of $w_{ij}^{\alpha}$ in Fig.~\ref{fig:L8Prod_h4}(a)
has no qualitative difference
from the case in the absence of $H_4$ [Fig.~\ref{fig:para_product}(c)].
However, we find that almost all products $w_{ij}^{1}w_{ij}^{2}$ are negative.
More impressively, as shown in Fig.~\ref{fig:L8Prod_h4}(b),
we find strong correlation between $w_{ij}^{1}$ and $w_{ij}^{2}$,
in sharp contrast to Fig.~\ref{fig:para_product}(d).
The symmetry breaking field $H_4 / T = 0.2$ completely orders the
couplings of the spins on the same lattice site to any hidden spin,
even though there is no apparent order in the coupling coefficients
for any individual color.
This provides further evidence to the validity of Eq.~(\ref{eqn:gamma})
as the order parameter for the product phase.

\begin{figure}[t]
\centering
\includegraphics[width=8.6cm]{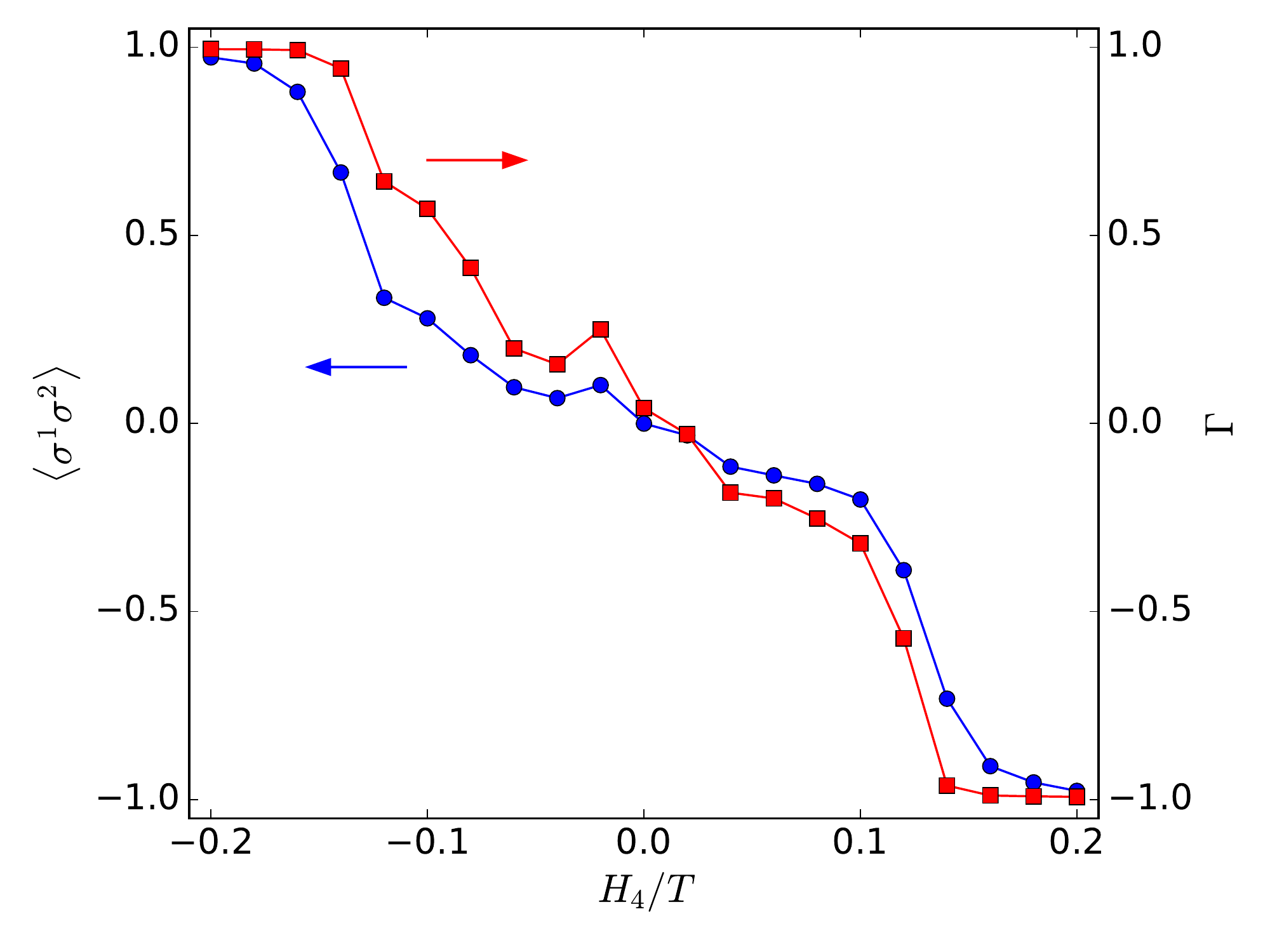}
\caption{Evolution of the machine-learning order parameters $\Gamma$ [Eq.~(\ref{eqn:gamma})]
and the conventional order parameter $\langle \sigma^1\sigma^2\rangle$
in the the two-color AT model on a 8$\times$8 square lattice for
$K_{2}/T=0.1$ and $K_{4}/T=1$ in the product phase
with an additional symmetry breaking field
$H_4 \sum_{i}\sigma _{i}^{1}\sigma _{i}^{2}$,
where $H_4 / T$  varies from -0.2 to 0.2.
}
\label{fig:Evolution}
\end{figure}

We further vary $H_4 / T$ from -0.2 to 0.2 and compare
the machine-learning motivated order parameter $\Gamma$ and
the conventional order parameter $\langle \sigma^1\sigma^2\rangle$
in Fig.~\ref{fig:Evolution}.
We find that when the $Z_2$ symmetry in $\langle \sigma^1\sigma^2\rangle$ is broken by
the external field that couples to the local order parameter,
the RBM study renders $\Gamma$ in {\em quantitative} agreement
with $\langle \sigma^1\sigma^2\rangle$.
Due to the sign function we choose, $\Gamma$ properly saturates to $\pm 1$
at sufficiently large $H_4 / T$, just as $\langle \sigma^1\sigma^2\rangle$ does.

\subsection{Generalization to the AT model with more color}
\label{sec:results.colors}

\begin{figure}
\centering
\includegraphics[width=8.6cm]{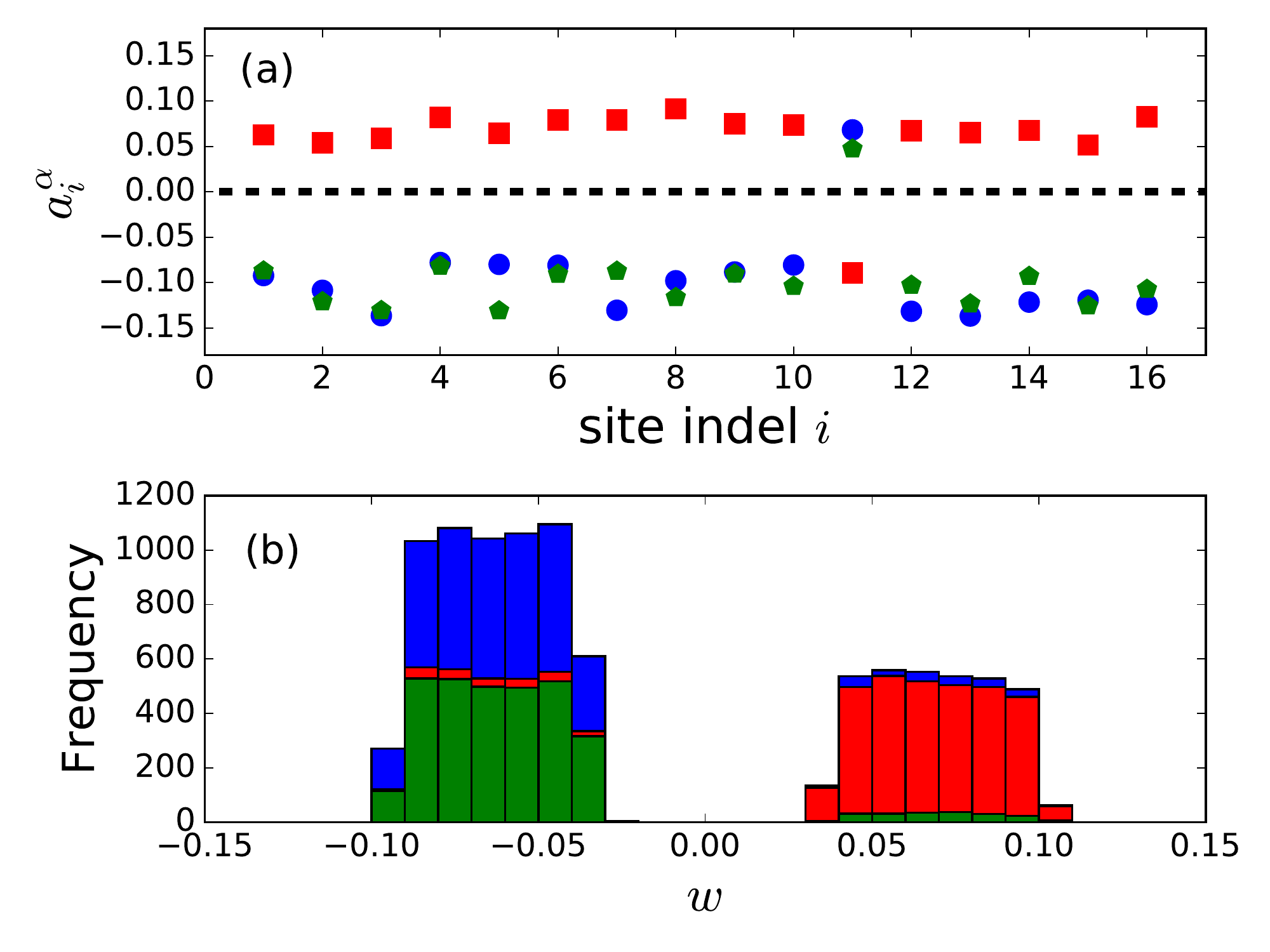}
% Here is how to import EPS art
\caption{RBM parameters obtained from the training of the three-color AT model
on a 4$\times$4 square lattice with $K_{2}/T=0.1$ and $K_{4}/T=1$ in the product phase.
The figure is color coded such that blue stands for color index $\alpha = 1$,
red for $\alpha = 2$, and green for $\alpha = 3$.
(a) The local fields for blue and green spins are of the same sign, which is opposite to
that of the red spins on the same site.
(b) The histogram of the coupling coefficients $w_{ij}^{\alpha }$ exhibits two lumps,
one on each side of the origin.
The left lump is larger and formed mostly by edges connecting
blue and green spins to hidden spins.
The right lump is smaller and formed mostly by edges connecting
red spins to hidden spins.
}
\label{fig:3Color-Product}
\end{figure}

We have now identified $\Gamma$, obtained from the products $w_{ij}^{1}w_{ij}^{2}$,
to be a legitimate order parameter in the RBM learning of the product phase
in the two-color AT model.
This prompts, however, interesting further questions. How do we generalize the results to
the $N_{c}$-color AT model for $N_{c}\geq 3$? Is the number of lumps
determined by the number of colors $N_{c}$? Can the product $w_{ij}^{\alpha
}w_{ij}^{\beta }$ (and $a_{i}^{\alpha }w_{ij}^{\alpha }$) continue to serve as
the indicators of the product phase? To answer these questions, we also
study the $N_{c}=3$ case as a self-consistent check. We plot the
local fields $a_{i}^{\alpha }$ and the histogram of the coupling
coefficients $w_{ij}^{\alpha }$ resulting from the RBM training of a 4$%
\times $4 lattice for $K_{2}/T=0.1$ and $K_{4}/T=1$ in the product phase in Fig.~%
\ref{fig:3Color-Product}, where all the training parameters are identical to
the two-color training case.
Ergodicity is chosen to be broken, again, in the data of spin configurations.
We find that $a_{i}^{\alpha }$ for color index $%
\alpha =1$ and $3$ have the positive sign and $a_{i}^{2}$ always assume the
opposite sign. Even though the sign for each color may not be the same, we
find $a_{i}^{1}a_{i}^{2}<0$, $a_{i}^{2}a_{i}^{3}<0$, and $%
a_{i}^{3}a_{i}^{1}>0$ for every lattice site index $i$. Correspondingly, we
find $w_{ij}^{1}w_{ij}^{2}<0$, $w_{ij}^{2}w_{ij}^{3}<0$, $%
w_{ij}^{3}w_{ij}^{1}>0$, and $a_{i}^{\alpha }w_{ij}^{\alpha }>0$. These
results, as we argued in the two-color case, are consistent with the product
order $\langle \sigma _{i}^{1}\sigma _{i}^{2}\rangle <0$, $\langle \sigma
_{i}^{2}\sigma _{i}^{3}\rangle <0$, and $\langle \sigma _{i}^{3}\sigma
_{i}^{1}\rangle >0$.
We emphasize that the signs depend on the sampling procedure
during which the ergodicity is broken,
but the same physics can be expected.

The combination of the training results in the product phase of the
two-color and the three-color AT models suggests that
deep in the product phase when ergodicity in the spin configurations is
completely broken, a generic $N_c$-color AT model features two lumps
in the histogram of $w_{ij}^{\alpha}$, one on each side of the origin.
The sign of either $w_{ij}^{\alpha}$ depends on the history of
training and the site index $i$, but the sign of the products
$w_{ij}^{\alpha} w_{ij}^{\beta}$ does not depend on the site index
$i$ or $j$ of the visible or hidden Ising spin.

\section{Optimal number of hidden neurons}
\label{sec:optimum_hidden}

In this section we discuss the choice of the number of hidden nodes,
which can help or hinder the understanding of the machine learning results.
Increasing the number of the hidden nodes can increase the representative
power of the RBM.
For example, in learning the thermodynamics of the Ising model,
Torlai and Melko~\cite{Torlai16} found that
the number affects the accuracy of the specific heat
when the system is at criticality.
Carleo and Troyer~\cite{Carleo17} studied the variational representation
of quantum states based on the RBM and found that the neural-network
state can achieve better accuracy when the number ratio of the hidden
nodes to visible nodes increases.
On the other hand, overfitting is also known to happen for too large a set
of parameters if the training data is redundant.~\cite{hintonRBM}

\begin{figure}[t]
\centering
\includegraphics[width=8.6cm]{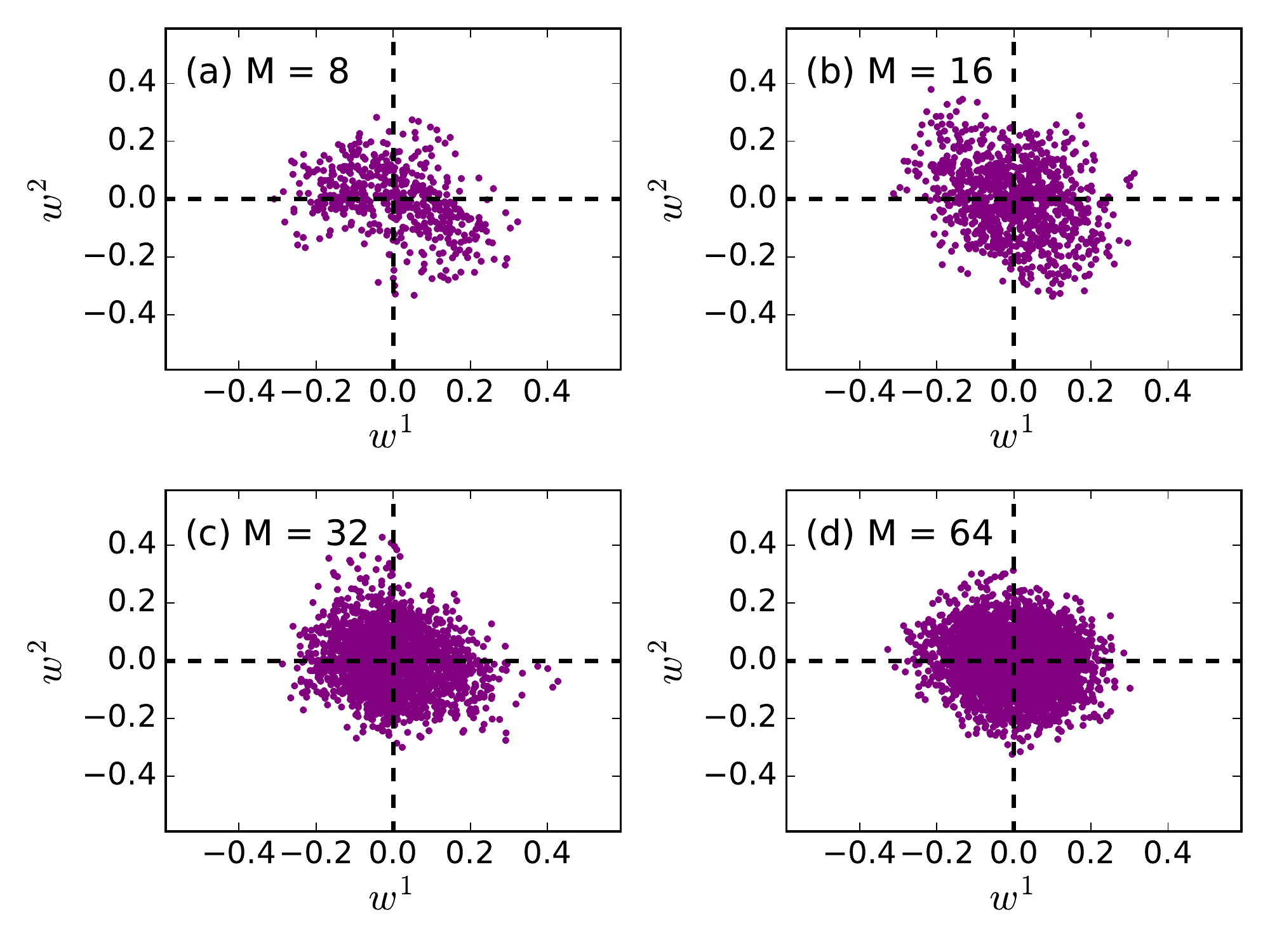}
\caption{Coupling coefficient $w_{ij}^{1}$ against $w_{ij}^{2}$
for various number of hidden nodes $M$ in the two-color AT model
with $L=8$ at $K_{2}/T=K_{4}/T=0.1$
in the paramagnetic phase.
For small $M$ in (a)-(c), fluctuation-led angular dependence is visible.
The angular dependence is negligible for $M = 64$.
}
\label{fig:L8Para_nhs}
\end{figure}

In the present unsupervised learning, we concentrate on the product order
of the spins on the same sites.
Two factors can obstruct our understanding.
First, in the paramagnetic phase, as well as in the product phase
when ergodicity is preserved, the distribution
of $w_{ij}^{\alpha}$ is expected to be color-blind,
i.e., independent of $\alpha$.
In Fig.~\ref{fig:L8Para_nhs} we plot $w_{ij}^{1}$ against $w_{ij}^{2}$
for an $L=8$ lattice in the paramagnetic phase.
For $M \leq 32$, some angular dependence of
$w_{ij}^{1}$ on $w_{ij}^{2}$ is visible, indicating that
the number of hidden nodes is not large enough.
The angular fluctuations can be neglected for $M = 64$,
which we identify as the smallest number of hidden nodes
that is required for $L = 8$, i.e., one hidden nodes per site.

\begin{figure}[t]
\centering
\includegraphics[width=8.6cm]{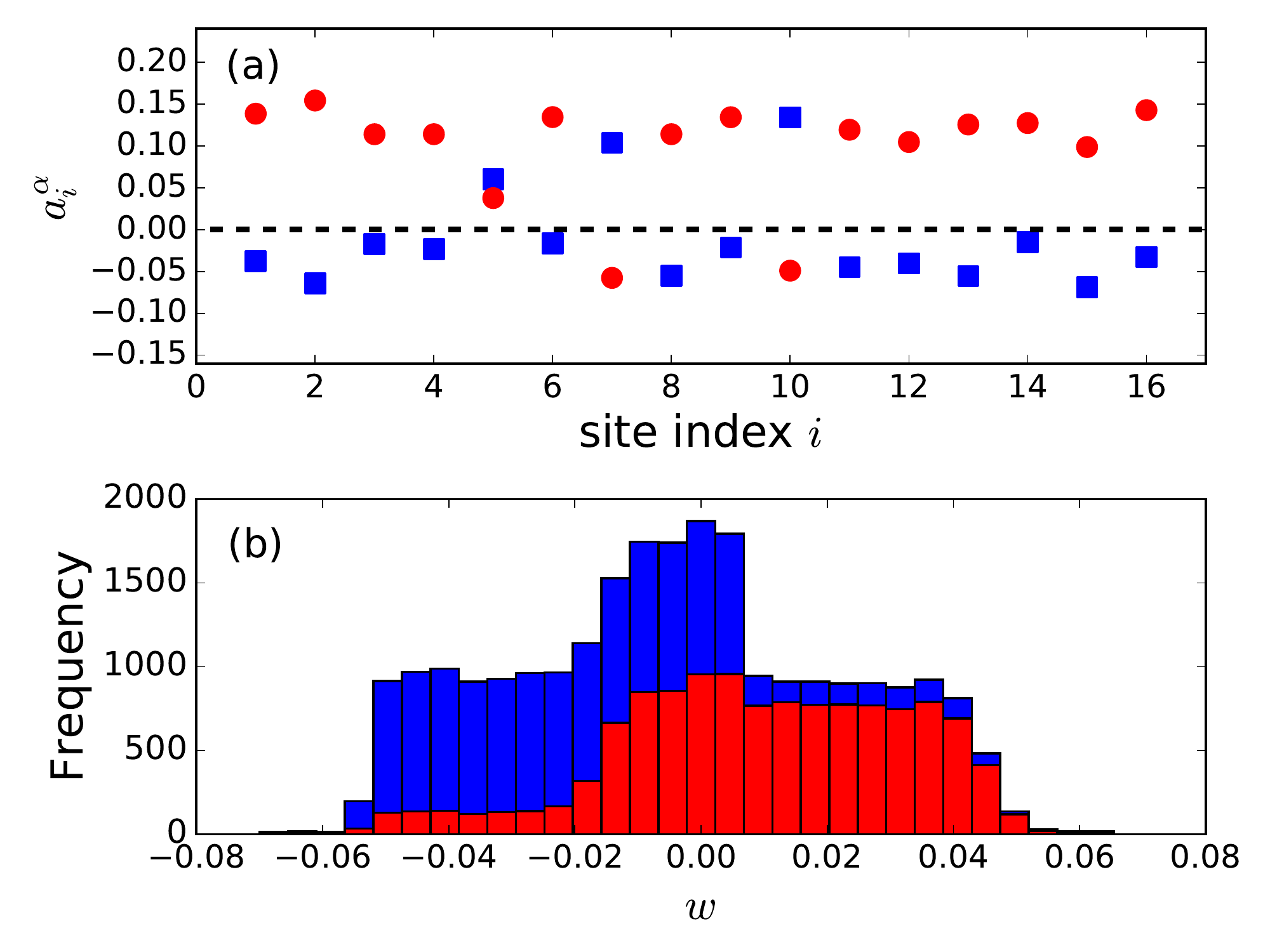}
\caption{
RBM parameters obtained from the training of the two-color AT model
on a 4$\times$4 square lattice with $K_{2}/T=0.1$ and $K_{4}/T=2$
in the product phase.
We choose the number of hidden nodes to be $M = 800$.
(a) The local fields $a_{i}^{1}$ and $a_{i}^{2}$ for the two-color Ising spins
on any site $i$ are no longer of opposite signs (see the violation on site 5).
(b) The histogram of the coupling coefficients $w_{ij}^{\alpha }$
is flat but with a bump around the origin.
The figure is color coded such that blue stands for color index
$\alpha = 1$ and red for $\alpha = 2$.}
\label{fig:L4_Prod_nh_800}
\end{figure}

The second factor arises because the polarization of the RBM parameters
and the subsequent two-lump structure in the product phase
is sensitive to the number of hidden nodes.
If the number of the hidden nodes is too large,
the majority of their weights can be rather small, leading to
a bump at the center of the distribution of $w_{ij}^{\alpha}$,
as illustrated in Fig.~\ref{fig:L4_Prod_nh_800}, in which
the number is chosen to be an unnecessarily large 800 for
$L = 4$.
This is consistent with fact that the product phase is a partially
disordered phase, so the presence of bump indicates that the
couplings between some visible spins are weak or negligible.
To clearly observe the two-lump structure, we find that
the number of the hidden nodes cannot exceed $300$
for $L = 4$, i.e., less than 10 hidden nodes
per visible spin, regardless of color.
We also note that $a_{i}^{1} a_{i}^{2} < 0$ is violated on the fifth site.

Combining the two factors, we choose the number of hidden nodes to
be 200 in the previous section, where we present our training results.

\section{Summary and Discussion}
\label{sec:discussion}

We have trained the RBM to study the 2D two-color AT model and
identified corresponding patterns in the RBM parameters
for ferromagnetic, paramagnetic, and product phases.
The ferromagnetic phase is characterized by a broad distribution of
interlayer couplings $w_{ij}^{\alpha}$ and polarized external
fields $a_{i}^{\alpha }$, which are consistent with the spontaneous
Z$_2$ symmetry breaking in the Ising spins.
The paramagnetic phase is characterized by a relatively narrower
distribution of the coupling coefficients $w_{ij}^{\alpha }$ with a zero mean
and unpolarized external field $a_{i}^{\alpha }$, which imply that
the Z$_2$ symmetry of the Ising spins are conserved.
The nontrivial product phase in the AT model is disordered
with regard to spins of a single color, i.e. $\langle \sigma
_{i}^{\alpha }\rangle =0$, but ordered with regard to the product of
the two spins on the same site,
i.e. $\langle \sigma_{i}^{\alpha }\sigma _{i}^{\beta }\rangle \neq 0$.
In the product phase, we find that the products $w_{ij}^1 w_{ij}^2$
are polarized and can be used to construct an order parameter
which quantitatively mimics $\langle \sigma_{i}^{\alpha }\sigma _{i}^{\beta }\rangle$.
The polarization of either $w_{ij}^{\alpha}$ or $a_i^{\alpha}$
depends on training and cannot serve as an indicator of
the product phase.
These results can be straightforwardly generalized to
the generic $N_c$-color AT model.

The RBM learning of the Boltzmann distribution is an unsupervised
learning. We only feed in the spin configurations by Monte Carlo
simulation. We do not provide any knowledge from our physical
understanding of the phases and corresponding phase transitions.
Through the distribution and polarization of the parameters or the
products of parameters, the RBM can distinguish disordered, partially
ordered, and ordered phases.
As demonstrated in the product phase, the information not only provides
hints on what order is established in the corresponding phase,
but also facilitates the construction of order parameter.
This is an attractive application of various machine learning schemes,
in particular, in poorly-studied models or
models without local order parameters,
as evident in other machine learning studies.~\cite{ohtsuki16}

The success in the identification of the product phase,
as well as of the ferromagnetic phase and the paramagnetic phase,
allure us to explore beyond training translationally invariant classical models.
One direction is to introduce disorder. The numerically obtained
RBM parameters do not respect translational invariance.
However, the sign of the parameters
(e.g., $a_i^{\alpha}$ in the ferromagnetic phase)
or that of the products of the parameters
(e.g., $w_{ij}^1 w_{ij}^2$  in the product phase)
respects translational invariance and
can be used to replace local order parameters
as the indicator of the corresponding phase.
It would be interesting to ask to what extent the results can be
generalized in the disordered AT model.
The disordered model also features a continuous phase transition rounded
by disorder from a first-order phase transition and the
emerging critical behavior is shown to be in the clean 2D
Ising universality class, accompanied by universal logarithmic
corrections.~\cite{Zhu15}
Whether it is possible to distinguish a continuous transition from a first-order one
in machine learning and how to identify the universality class of a continuous
transition are interesting questions.
Another interesting direction is to learn topological quantum systems,
whose topological properties are robust against disorder.
The Kitaev model is one such example, which has simple and
translationally invariant solutions when one represents
the quantum many-body ground state by an RBM.~\cite{Deng16}
To what extent numerical training can identify the topology
in such systems is of great interest.~\cite{YiZhang16}

\section{Acknowledgements}

This work is supported by the National Basic Research Program of China
through Project No. 2015CB921101 and the National Natural Science Foundation
of China through Grant No. 11674282.

\end{document}